\documentclass[sigconf,natbib=false,nonacm]{acmart}

\setcopyright{acmlicensed}
\setcctype{by-nc}
\copyrightyear{2024}
\acmYear{2024}

\RequirePackage[
datamodel=acmdatamodel,
style=acmnumeric,
backend=biber,
]{biblatex}

\usepackage{hyperref}
\usepackage{todonotes}

\begin{document}

\title[Introducing the NACSOS-nexus platform]{NACSOS-nexus: NLP Assisted Classification, Synthesis and Online Screening with New and EXtended Usage Scenarios}

\author{Tim Repke}
\email{repke@mcc-berlin.net}
\orcid{0000-0001-9661-6325}

\author{Max Callaghan}
\email{callaghan@mcc-berlin.net}
\orcid{0000-0001-8292-8758}
\affiliation{%
	\institution{Mercator Research Institute on Global Commons and Climate Change (MCC), Berlin, Germany}
	\city{}
	\country{}
}
\affiliation{%
	\institution{Potsdam Institute for Climate Impact Research (PIK), Germany}
	\city{}
	\country{}
}

\renewcommand{\shortauthors}{Repke et al.}

\begin{abstract}
NACSOS is a web-based platform for curating data used in systematic maps.
It contains several (experimental) features that aid the evidence synthesis process from finding and ingesting primary data (mainly scientific publications), basic search and exploration thereof, but mainly the handling of managing the manual and automated annotations.
The platform supports prioritised screening algorithms and is the first to fully implement statistical stopping criteria.
Annotations by multiple coders can be resolved and customisable quality metrics are computed on-the-fly.
In its current state, the annotations are performed on document level.
The ecosystem around NACSOS offers packages for accessing the underlying database and practical utility functions that have proven useful in a multitude of projects.
Further, it provides the backbone of living maps, review ecosystems, and our public literature hub for sharing high-quality curated corpora.
\end{abstract}

%



\maketitle

\section{Introduction}
The exponentially growing number of publications published each year poses a significant challenge for authors of synthesis reports, such as those by the IPCC.\footnote{The Intergovernmental Panel on Climate Change (IPCC) is a scientific body that assesses the latest scientific, technical, and socio-economic information produced worldwide relevant to the understanding of climate change, its potential impacts, and options for adaptation and mitigation.}
Therefore, systematic maps and reviews play a vital role in aggregating primary evidence that condense the vast amount of available information.

The methodology for synthesising scientific evidence was continually refined over the last decades~\cite{bates2007systematic,tsafnat2014systematic,james2016methodology,nakagawa2019research}.
There are two typical ``kinds'' of synthesis projects:
\emph{Systematic maps} with the goal of providing a comprehensive overview of a specific research area or identify sparse areas and gaps in primary evidence~\cite{saran2018evidence} and \emph{systematic reviews}, that dive deeper into particular clusters of evidence while collecting and normalising particular indicators to answer narrow, pre-defined research questions.
The core steps include 
1) \emph{searching and retrieving evidence} (typically using iteratively revised boolean queries in bibliographic databases);
2) \emph{screening evidence} (reviewing relevance of retrieved records and including/excluding them following clear guidelines);
3) \emph{information extraction} (detailed annotation following a codebook and collection of meta-data);
4) \emph{critical appraisal} (assessing study quality for internal and external validity); and
5) \emph{evidence synthesis} (compiling all gathered facts into key insights).

Several software solutions have been built in order to support authors of (systematic) evidence synthesis projects, such as ASReview~\cite{van2021open}, EPPI-reviewer~\cite{thomas2010eppi,thomas2023eppi}, covidence~\cite{covidence}, NACSOS~\cite{max_callaghan_2020_4121526}, and many others.
Although core features are similar, they are each tailored to specific use-cases, needs and scientific communities.
To this end, we created and are continually developing \emph{NACSOS-nexus}---a complete architectural redesign and rewrite of the original \emph{NACSOS} platform---in order to address the specific requirements that arise in the APSIS (applied sustainability science) group at the MCC Berlin (soon part of PIK Potsdam).
The entire project is fully open source and can be hosted anywhere.\footnote{\url{https://apsis.mcc-berlin.net/nacsos-docs/}\label{docs}}
Many collaboration partners use the primary instance maintained at the MCC Berlin, which hosts 5.6M documents (of which 50k are manually annotated) across more than 40 research projects and is used by 120 scientists and student researchers from over 40 institutions.\footnote{Numbers retrieved on May 6th, 2024.}

The platform is mainly built for conducting systematic maps.
It is well suited for the specific workflow from ingesting (bibliographic) data from most popular sources, screening and coding documents (on document-level), and managing assignments and annotations by multiple coder, resolving disagreements, and keeping track of progress and data quality metrics.
It is, at this point, not well suited for query development or the specific needs of systematic reviews or qualitative coding tasks.
The platform is most often used for projects where the available evidence is in the order of tens- to hundreds of thousands of documents that require machine-learning assisted approaches with statistical criteria to estimate confidence in completeness and quantifying uncertainties.

For a more complete description of all features and details, we refer readers to the documentation\footnotemark[2] or to get in contact with us, as this is beyond the scope of this work. 
Please also note, that this article is not meant to be a complete or up-to-date description of the platform.
It is meant to highlight key features and concepts as a reference to cite in your survey, map, review, or other article that uses the platform.

In the remainder of this article, we provide an overview of the architecture and concepts; specifics about some of the key features, particularly data ingestion, prioritised screening, and organising annotations; and share the vision of the extended ecosystem being developed around the platform.

\section{Architecture overview}
In this section, we introduce the main concepts and terminologies including their relation to one another.
Further, we briefly review the technologies used by \emph{NACSOS-nexus}.

\paragraph{Concepts and data structure.}
Almost all data captured by the database relates to an \texttt{item}, which is ``unique'' within each project.
Since the platform can handle arbitrary text documents, an \texttt{item} is the most generic form and the uniqueness, available meta-data, and way of displaying that data is different for scientific publications (see below for details), tweets, patents, grants, or news articles.
The uniqueness constraint is not strictly enforced, but assumed by many downstream processes.
During the \texttt{import} of a new data source, the new data is merged into the existing corpus of a project.
For scientific articles, the platform keeps track of alternative variations of detected duplicates, since many scientific databases are not consistent or may have different meta-data for the same underlying publication.
In particular, we treat different versions of the same publication as a single item.
Although one could argue to keep them separate, our use-case is typically only concerned with the ``information'' or piece of evidence, which would not change between a pre-print and its published version.
We keep track of all links between items and imports, which later allows us to analyse overlaps between different data sources.

All \texttt{annotation}s are linked to an \texttt{assignment} which are grouped in \texttt{assignment\_scopes}.
In this way, we are able to create batches of documents to be annotated next and assign specific users to these units of work.
Typically, the assignments are random (to create reference datasets) or ranked by predicted relevance according to coding guidelines defined in the respective protocol.
Each item is assigned to multiple users to retain gold-standard annotations where disagreements are discussed and resolved in after each batch of assignments.
Users do not see the other's annotation before beginning the resolution process.
During the resolution, coding guidelines are refined.

An \texttt{assignment} forms a triplet that links an item, user, and \texttt{annotation\_scheme}.
This scheme defines which \texttt{label}s users can (and have to) annotate for each document.
A label can be binary (e.g. relevance yes/no), single-choice (choose exactly one from a list of options), multi-choice (choose multiple from a list of options), or free-text (e.g. for comments or extracted information).
Labels can be nested and may also be repeatable (order-sensitive, e.g. for annotating the primary and secondary value for a label).

In much the same way, the platform uses \texttt{bot\_annotations} with meta-data scopes to keep track of automatically generated annotations or system-internal annotations such as resolved labels.


\paragraph{Database and backend.}
All data is stored in a PostgreSQL database across 25 tables and currently holds 55GB (52M tuples).
The database schema is reflected in SQLAlchemy2\footnote{\url{https://docs.sqlalchemy.org/}} and migrations are handled by alembic.\footnote{\url{https://alembic.sqlalchemy.org/}}

For convenience, we develop and maintain the Python package \texttt{nacsos\_data}\footnote{\url{https://gitlab.pik-potsdam.de/mcc-apsis/nacsos/nacsos-data}} for interacting with the database.
It contains many commonly used CRUD functions for creating, reading, updating, or deleting data.
The package also contains a library of readers for several file-types (e.g. Scopus csv exports), translators into the \emph{NACSOS-nexus} data format, retrieving data via APIs, deduplication algorithms, as well as other useful utility functions for handling and analysing data.
The package is fully typed using pydantic,\footnote{\url{https://docs.pydantic.dev/}} which allows for automated testing, API generation, and easier access for other developers as the expected inputs and outputs are clearly defined (beyond documentation and code comments).

The server backend, \texttt{nacsos-core},\footnote{\url{https://gitlab.pik-potsdam.de/mcc-apsis/nacsos/nacsos-core}} uses \texttt{nacsos\_data} package and exposes most of the package's functionality through FastAPI.\footnote{\url{https://fastapi.tiangolo.com/}}
This REST-API also implements additional monitoring, administration, and authentication endpoints.

\paragraph{Background tasks.}
Many features of the platform require long-running tasks, for example data ingestion, data analysis, or training and applying classifiers or other machine learning models.
Such tasks are triggered via an event system within the backend or via the API.
Each new task is pushed into a queue and picked up by separate worker processes using dramatiq,\footnote{\url{https://dramatiq.io/}} a modern and more flexible and extensible distributed task queueing library challenging to replace celery.

The \texttt{nacsos-core} backend also wraps the native dramatiq tasks to keep track the task status in the database and manage artefacts, such as use uploads, processing logs, and generated figures or other files.
Further, it may inject dependencies or re-organise queue prioritisation based on the available (and estimated) runtime resources.
In this way, a user may schedule a hyper-parameter sweep for testing different settings for a topic model, while the queue manager would intermittently prioritise single tasks by other users as to not keep them waiting too long, or prevent simultaneous execution of two resource-intense tasks.

We built the environment and parameter injection in such a way, that all functions can be re-used outside of the server context in custom scripts.

\paragraph{Web-based frontend.}
The vast majority of users interacts with the platform through the web interface.\footnote{\url{https://gitlab.pik-potsdam.de/mcc-apsis/nacsos/nacsos-web}}
The interface is implemented in the Vue3.js\footnote{\url{https://vuejs.org/}} framework using Typescript.
Thanks to the fully-typed data library and backend, FastAPI can generate an API specification that we use to generate a typed API client.
This allows automated type checks and helps us to quickly identify and refactor relevant code, should an API change.

Depending on the user's permission per project, they will only have access to respective features, such as configuring and managing imports, exploring and querying the project corpus, or managing annotations and assignments.
The most time spent on the platform is in the annotation views.
To this end, we pay most attention in delivering a smooth and efficient user experience.
Once logged in, users can track their open and fulfilled assignments.

\begin{figure*}[h]
	\centering
	{%
		\setlength{\fboxsep}{0pt}%
		\setlength{\fboxrule}{1pt}%
		\fbox{%
			\includegraphics[clip, trim=0 2.6cm 0 0, width=\textwidth]{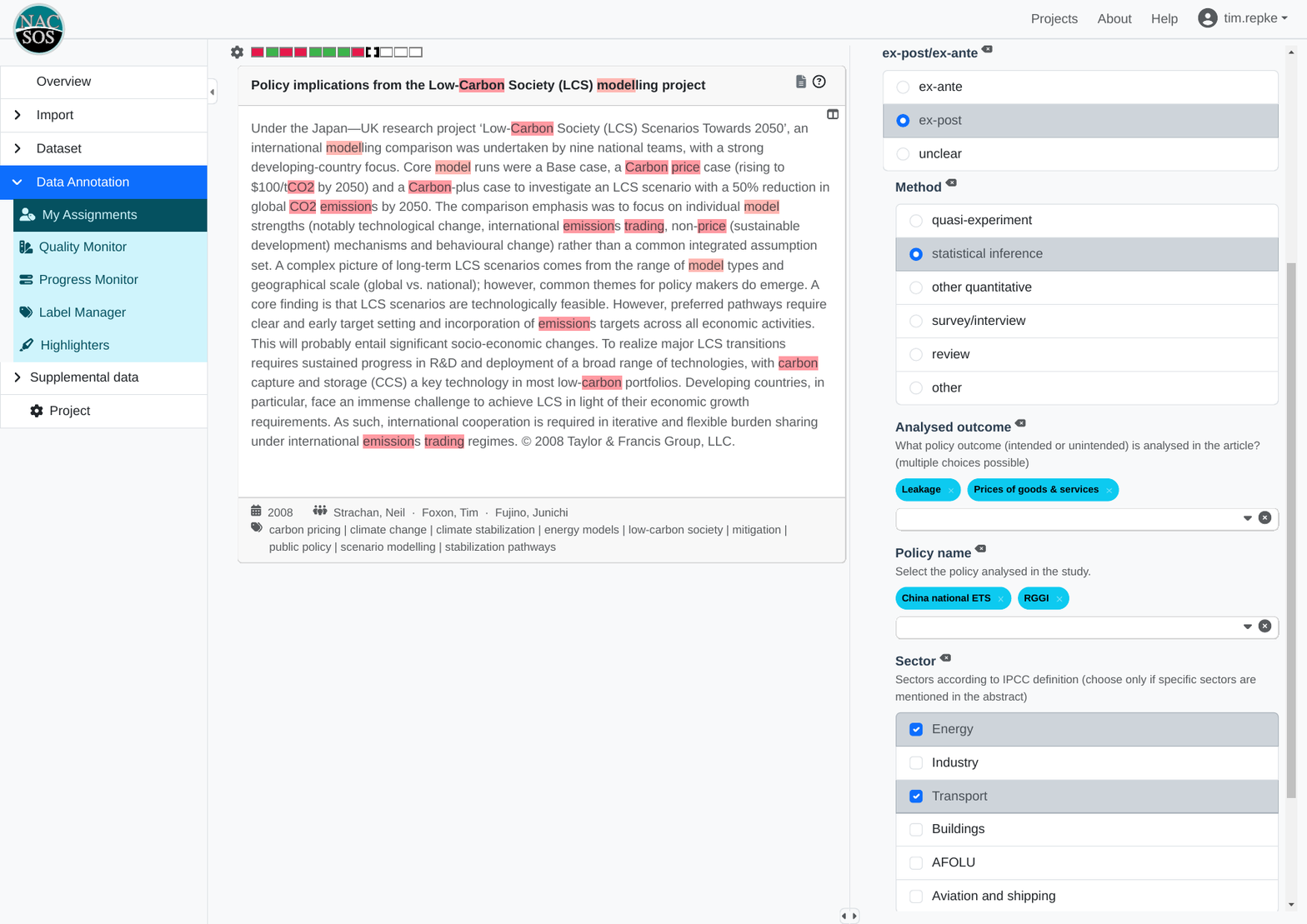}%
		}%
	}%
	\caption{Item annotation view; document is shown in the middle, the annotation scheme on the right. This example shows a nested scheme (some labels only show up when another condition is met) with several single- and multi-choice labels.}
	\label{fig:annotation}
\end{figure*}

In the item annotation view (see Figure~\ref{fig:annotation}), users see the respective document with all available meta-data, a progress bar for the active assignment scope, and the annotation scheme.
The colours in the progress bar are configurable by assignment status or any other label.
This feature turned out to be very practical for jumping back to other already coded documents for comparison, which also improves annotation consistency.
Most annotations can be performed ergonomically and efficiently using keyboard shortcuts.
Furthermore, the item annotation view is fully responsive, which is great for performing simple annotations on the phone while commuting or on the couch.

\section{Data ingestion}
\emph{NACSOS-nexus} natively supports several types of text documents, such as scientific publications, tweets, LexisNexis articles, and patents.
Furthermore, it can handle any generic document and will store and display JSON-compatible meta-data.
When needed, any other specific data types can be implemented fairly easily.

\paragraph{Data sources.}
At this point, we support a range of different data sources and are continuously adding more formats.
Most importantly, we can import tweets directly from the Twitter Developer API, DimensionsAI, Scopus CSV exports, Web of Science RIS exports, and the LexisNexis API.
Most of these imports can be configured directly in the web interface.

Further, more specific import modes, particularly in frequently (automatically) updated projects, such as for living evidence synthesis projects, are under development and currently handled via external pipelines.

\paragraph{OpenAlex.}
We host a snapshot of the open source database of 250M publications OpenAlex~\cite{priem2022openalex} using solr (for full-text search) and a dedicated PostgreSQL database (for references, full meta-data support, and in-depth scientometric analyses).
We implemented an automated pipeline for ingesting the latest snapshots via GitLab.

This self-hosted snapshot enables several advanced querying and analytical features.
Having access to the full vocabulary and database index has already been very revealing.
For example, translating wildcard keywords can be critical.
In climate-related literature searches, the wildcard ``clim*'' also returns many unintended results such as variations of ``climb'' or ``climax''.

Furthermore, the use of an open database provides full transparency about the original data sources.
Paid data providers are not only very costly, but also do not return the same results for the same query when using access from different institutional subscriptions.
We can also run comparative queries with ease and provide relative counts to normalise overall publication count trends (e.g. proportion of annual articles of overall literature) to better estimate the actual growth of a field and not just the overall growth in publications.

\paragraph{Deduplicating academic items.}
For most use-cases, it is critical to operate on a set of distinct documents.
For example, manually annotating data is very resource- and time-intensive.
Coding the same documents multiple times would be wasteful.
Also, any downstream analysis such as topic models or raw publication counts for a particular category will degrade in quality with duplicates.

To this end, we implemented multi-stage deduplication strategies that are specific to each type of data.
In this paragraph, we only focus on academic items as an example.

Scientific publications are usually assigned a DOI (document object identifier) that can be used to uniquely identify specific articles.
However, at scale, we encounter quality issues where multiple DOIs point to the same article or clearly different articles are assigned the same DOI in the source database.
Other source specific identifiers (e.g. Web of Science ID, Scopus ID, OpenAlex ID,...) shift the responsibility of proper deduplication to the data provider.
Again, at scale, we learned that these identifiers cannot be fully trusted to enforce uniqueness constrains.
Aside from data quality issues, this can also be attributed to the definition of ``uniqueness'', which usually considers versions (e.g. preprint and final publication) as different articles.
For our use-case, these should be considered duplicates and therefore merged into one item.

We consider anything where title and abstract (when both are available) are overwhelmingly similar as duplicates in accordance with the above mentioned definition.
First, we strip any stop-words and characters other than a-z from the concatenated, lower-cased title and abstract.
We also remove other common terms, such as ``abstract'', which are frequently appearing artefacts in academic databases.
From this, we transform the texts into sparse binary token frequency vectors that are indexed in an approximate nearest neighbour index using pynndescent.\footnote{\url{https://github.com/lmcinnes/pynndescent}}
Such an index allows us to find similar documents very fast with at a reasonable memory footprint even for larger datasets.
The similarity between two texts is computed using the Jaccard index, which is defined as
$$
J(A,B) = {{|A \cap B|}\over{|A \cup B|}},
$$
where A is the set of tokens in the one text and B the set of tokens of the other text.
In other words, it is the ratio between tokens that appear in both texts and all tokens used by either text.
For example, a near-duplicate text with ten words, where the other candidate is missing one of those, would result in a score of 90\%.
In practice, we usually require a minimum length of 10 tokens for title and abstract with an overlap (concatenated text) of at least 95\% to find duplicates.

For other cases, where abstract is missing, we employ a different strategy based on title and publication years.
First we create an index of ``title slugs'', which is a radically trimmed title containing only lower-case letters (even excluding spaces).
For all matches, we check the title for numbers that indicate a version, chapter number, or publication year (those might be false-positive matches).
Further, we check the publication year, but allow for a difference of $\pm$1 year to account for data quality issues or different article versions.
Experiments using the author lists have proven to be very complex due to differing and sometimes inconsistent formatting conventions as well as different cut-offs (some data sources do not include all authors for longer author lists).

Once we identified duplicates, we update auxiliary indices for duplicate detection and fuse the new item into the database.
In the fusion process, we try to complete the existing item when some meta-data is missing.
When we encounter different values for a field (e.g. different DOIs), we keep track of all variants in a dedicated table.
In this way, we are able to reconstruct other versions and values at a later time and retain the data lineage.

\section{Prioritised annotation}
%
%
%
\emph{NACSOS-nexus} contains tools to assist in the process of machine-learning-prioritised screening.
It uses machine learning to continually rank all unlabelled documents by ``relevance'' (which studies to include in the map or review) and assign the next most relevant labels to coders for annotation.
In this way---compared to screening the entire dataset by hand---the effort required to screen documents can be reduced significantly.
This process is also sometimes referred to as ``active learning''~\cite{omara-eves_using_2015}, but actually varies from the usual definition of AL, in that the goal is to see all relevant documents by the end of the process, not just to label the least number of documents that are most impactful for the model to learn how to distinguish which studies to include or exclude based on their title and abstract.

In practice, the following process is followed:
\begin{enumerate}
	\item Annotate a random selection of documents
	\item Train a classifier using the annotated data, to predict the potential relevance of unseen documents
	\item Assign the N documents with the highest predicted relevance to be screened in descending order
	\item Repeat steps (2) and (3) until it is estimated that enough relevant documents have been identified
\end{enumerate}

Step 4 requires methods to decide when ``enough''---ideally all---relevant documents have been identified.
\emph{Nacsos-nexus} provides an interface that allows users to make decisions on when to stop screening based on a statistical stopping criterion~\cite{callaghan2020statistical}.
Since we are convinced that stopping criteria are essential for confident decisions, we also provide this feature as dedicated packages in Python and R, as well as a web-based tool.\footnote{\url{https://apsis.mcc-berlin.net/project/buscar/}}

The underlying classification (or regression) model used for ranking may vary between projects.
Even traditional logistic regression models on bag-of-word vectors can sometimes provide similar performance as state-of-the-art transformer models at a fraction of the computational cost.
We have a set of scripts for ``best practices'' and are working on a unified toolkit that automatically tests a wide array of models and hyper-parameters to find the best solution for each use-case.

\section{Monitoring annotation quality}
With multiple annotators working on the same project, it is important to keep track of the progress and monitor the annotation quality.
At the same time, we feel that it is critical not to develop tools that indirectly foster unhealthy working environments by keeping track of time spent, pressuring for quotas, or quantifying which annotator might disagree more often with others.
Although these might be helpful tools to estimate time needed for future assignments, we learned that direct communication within the author and annotator team is more open and flexible and less stress-inducing environment.

An integral part of the annotation process is the label resolution view (see Figure~\ref{fig:resolution} in the appendix).
It provides an overview of all labels (and their values) for an assignment scope.
After all coders finished their assignments, the team would usually go through the annotations using this view and discuss disagreements and consolidate labels accordingly.

Other views include quality metrics at varying levels of detail.
These include Pearson correlation, Kendall's tau, Cohen's kappa, overlap (of multi-labels), precision/recall, agreement (for a configurable rule across multiple labels), and many others.
This monitoring if often used to make strategic decision when to move from triple- to double-coding or even partial double-coding, where only a proportion of items is annotated by more than one coder.

Additionally, we implement an annotation progress monitor.
This is used in larger projects that use prioritised screening.
In these cases, it is not clear when to stop annotating (see above).
Our statistic stopping criteria can be used to make an informed decision and keep track of confidence scores for different freely configurable ``inclusion'' rules.

\section{Querying items}
Retrieving specific data for particular use-cases is an important aspect of this platform.
For example, users like to define specific and sometimes complex rules for which documents to annotate next, include in an analysis, or to export.
It turned out, that implementing all potential filters (and their combination logic) quickly becomes infeasible, results in hard-to-understand interface components, and is hard to maintain as new features are added.
To this end, we created the \emph{NACSOS Query Langual (NQL)}.
It is strongly inspired by SQL and is meant to be simple, yet expressive enough to expand to capture very specific rules.

Using NQL, users can query for any annotation or set of annotations within a project, search item meta-data, and nest and combine those rules into complex sub-queries.
NQL can be used anywhere on the platform where data can be selected or filtered.
It has also proven to be useful in other applications outside of the user interface, as it is more concise and easier to use than manually writing complex database queries.
After some initial iterations, the grammar has not evolved further in some time.
The current version is using nearley\footnote{\url{https://nearley.js.org/}} in the frontend to convert NQL into json (to be handled by the backend).
In future versions, we plan to build an interactive NQL editor that makes suggestions based on the current parse tree context.

\section{Beyond the platform}
The \emph{NACSOS-nexus} platform is just the starting point for many of our evidence synthesis projects.
While there are some common analyses and figures that can be re-used across multiple project, further analysis and processing of the annotated data is usually done outside of the platform (or fed back into the platform).
As mentioned before, the \texttt{nacsos\_data} package contains a library of utility functions for accessing and processing the data stored in the underlying database.
We are also adding generalisable processing pipelines where possible, so that future projects can benefit from it.

We belief that users should retain full ``ownership'' of their data.
Thus, the platform offers a full export of all data collected within a project.
Exports can also be constrained to an NQL query, which is helpful for quick sharing of specific subsets.

Transparency and reproducibility are important cornerstones in science.
To this end, we are developing a dedicated literature hub---an interactive platform where anyone can get free access to explore the full original annotated corpus of literature used in a systematic map or review.
This is also a valuable contribution to the research community, as they get full access to curated data that they can use in their research project, particularly reviews of dedicated sub-topics or evidence gaps.
In this way, we are fostering a community of reviews and follow-up projects to benefit from the huge mapping efforts conducted in many of our projects. 

Furthermore, as we are using machine learning and automated pipelines, we are working towards a living evidence ecosystem.
New publications that match the search query of a project can be automatically classified or assigned to coders for annotation.
Once the new data is enriched, it can be integrated into the existing curated corpus and shared on the literature hub.
This ensures that any user of the synthesised evidence, such as other researchers or policy makers, always have access to the most recent relevant evidence when they need it.

\section{Conclusion}
The original \emph{NACSOS} platform has been proven useful in numerous projects.
The new \emph{NACSOS-nexus} platform that we redesigned from the ground up opened countless possibilities for future developments and novel features.
We are planning to develop stand-alone packages for generalisable features, such as priotisation, stopping criteria, or uncertainty estimation as part of an ecosystem around this project.
We invite developers of other similar platforms to challenge our methodology and validation strategies, contribute to the ecosystem, or integrate these features into their platform.
In this way, no matter which solution a user prefers, they have access to leading methods to conduct their studies following the latest standards and best practices.

In the near future, \emph{NACSOS-nexus} will provide an accessible pool of re-usable common analyses and figure generators, fully-customisable suite of classifiers to automatically choose the best one for the use-case, task-specific annotation interfaces for more efficient labelling (e.g. validating AI-suggested labels or automatically extracted data), and in-text (aka full-text) annotations.
Furthermore, we are working on thoroughly evaluating all critical automation steps of the platform, such as differences between data source coverage and quality (and the resulting impact on retrieving documents using queries), classifiers (and potential for biases or missing entire clusters), or estimating uncertainties when quantifying items using automated classifiers with varying precision/recall scores.
\newpage
\begin{acks}
This project received funding from the European Union's Horizon 2020 research and innovation programme under the European Research Council (ERC) in the GENIE project (grant no. 951542-GENIE-ERC-2020-SyG) and from the German Federal Ministry of Education and Research under the ARIADNE (grant no. 03SFK5J0) and CDR-SynTra (grant no. 01LS2101F) projects.
The content of this deliverable does not reflect the official opinion of the European Union nor German government.
Responsibility for the information and views expressed herein lies entirely with the author(s).
\end{acks}


\printbibliography

\appendix

\begin{figure*}[htb!]
	\centering
	{%
		\setlength{\fboxsep}{0pt}%
		\setlength{\fboxrule}{1pt}%
		\fbox{%
			\includegraphics[clip, trim=0 8cm 0 0, width=\textwidth]{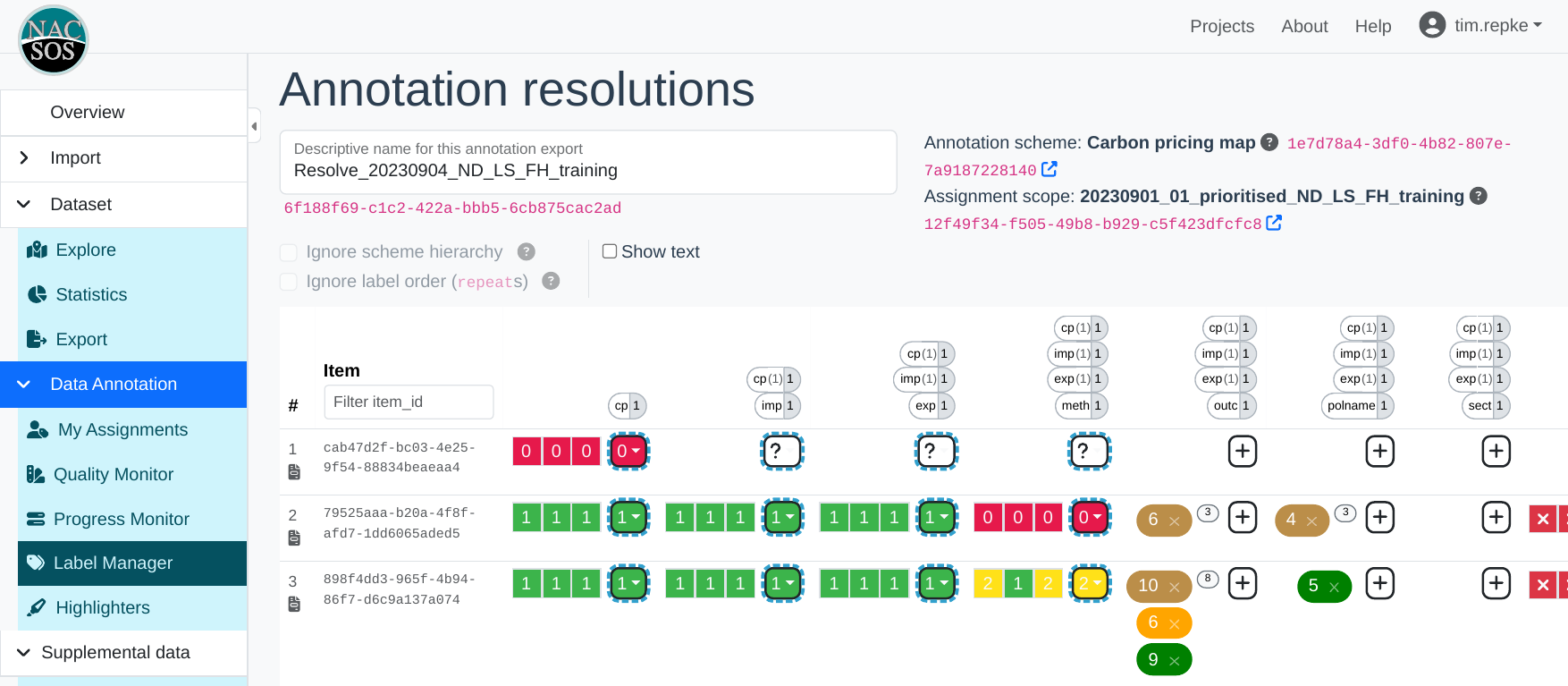}%
		}%
	}%
	\caption{Label resolution view. Rows contain annotations for this item, columns are (nested) labels. Rectangle boxes are each annotator's labels, rounded boxes are resolved labels. Hovering any field reveals additional details (such as human-readable label or value, name of annotators, or document information).}
	\label{fig:resolution}
\end{figure*}

%
%
%

\end{document}